\documentclass[aps,twocolumn,prb,showpacs,balancelastpage,amssymb,groupedaddress]{revtex4}
\usepackage{epsfig}
\usepackage{bm}

\begin{document}

\title{ Local and nonlocal entanglement for quasiparticle pairs induced by Andreev reflection }

\author{Zhao Yang Zeng$^{1,2,3}$}
\email{zyzeng@jxnu.edu.cn}

\author{ Liling Zhou$^{1,4}$}

\author{ Jongbae Hong$^2$}

\author{ Baowen Li$^{3,5,6}$}

\affiliation { $^1$Department of Physics, Jangxi Normal University, Nanchang 330027, China\\
$^2$School of Physics $\&$ Center for Theoretical Physics, Seoul National University, Seoul 151-747, Korea\\
$^3$ Department of Physics and Centre for Computational Science and Engineering, National University of Singapore, 117542, Singapore\\
 $^4$Department of Physics, Hunan Normal University, Changsha, 410081, China\\
 $^5$Laboratory of Modern Acoustics and Institute of Acoustics, Nanjing University, Nanjing 210093, China\\
 $^6$ NUS Graduate School for Integrative Sciences and
Engineering, Singapore 117597,  Singapore }
\date{\today}
\begin{abstract}

We investigate local and nonlocal entanglement of particle pairs
induced by direct and crossed Andreev reflections at the
interfaces between a superconductor and two normal conductors. It
is shown theoretically that both local and nonlocal entanglement
can be quantified by concurrence and detected from the violation
of a Bell inequality of spin current correlators, which are
determined only by normal reflection and Andreev reflection
eigenvalues. There exists a one-to-one correspondence between the
concurrence and the maximal Bell-CHSH parameter in the tunneling
limit.
\end{abstract}
\pacs{73.50.Td,74.45.+c, 03.67.Mn, 03.65.Ud \\[-5mm]}
\maketitle

      Quantum entanglement is a peculiar nonlocal correlation between distant
   quantum mechanical systems\cite{Terhal}. It plays a key role in the emerging tasks of
   quantum information, such as quantum cryptography\cite{Ekert}, quantum dense coding\cite{Bennet1}, quantum
   teleportation\cite{Bennet2},  quantum error correction\cite{Shor}, and quantum
   computation\cite{Bouwmeester}. Entanglement is also critical for enabling a quantum computer to perform some computational tasks
   exponentially faster than any classical
   one\cite{Jozsa}.
    This kind of  nonclassical
   puzzling correlation has been demonstrated experimentally for
   photons\cite{Aspect} and atoms\cite{Hagley}. However, it
   remains a challenging task for experimentalists to demonstrate
   electronic entanglement in solids.

   There is growing interest in controllable entanglement between electrons in solids, especially in
   mesoscopic systems. A variety  of schemes have been proposed,
   focusing on the creation,  manipulation, and/or detection of spin entanglement of
   electron pairs by using either quantum dots\cite{Burkard} or
   superconductors\cite{Lesovik, Recher}.  A mesoscopic
   normal-conductor-superconductor(NS)
   device has been designed for the creation and detection of orbital entanglement between electron
   pairs\cite{Samuelsson}. All of the proposals with hybrid
   NS structures only allow for
   ideal spin-independent tunneling at the interface between the normal
   conductor and the superconductor. The degree of entanglement is still not, to the best of our knowledge,
   quantified for electron pairs emitted from a
   superconductor. A novel idea for electron-hole entanglement produced in
   tunneling events in normal conductors has been recently
   put forward by Beenakker et al.\cite{Beenakker}.  Concurrence\cite{Wootters} and the Bell-CHSH
   parameters\cite{Bell}
   for entangled electron-hole pairs  have also been analyzed  based
   on the scattering matrix theory.

   In this work, we study local and nonlocal entanglement
   between pairs of quasiparticles(electrons and holes), which are created in Andreev
   reflection processes in hybrid normal-conductor-superconductor systems. As an example,
   we consider a structure with two normal conductors
   connected to a common superconductor via tunnel barriers\cite{Russo}, as
   depicted in Fig. $1$. The separation $d$ between these two normal
   conductors should be comparable to the superconducting coherence
   length $\xi$, in order to guarantee the occurrence of the crossed Andreev reflection process\cite{Deutscher}.
   One can expect entangled pairs of either electrons or holes depending on the sign of voltage applied
   to one of the two normal conductors. If a negative voltage
   $\mp eV$ is applied to the left conductor, an incident spin-up(down)
   hole(electron) in the left  conductor
   can be Andreev reflected as a spin-down(up) electron(hole) either  $(i)$ at the
   left conductor, or, $(ii)$ at the right conductor.
   The former  process is termed as  direct Andreev reflection, and the latter is  crossed Andreev reflection
   \cite{Deutscher}. Since the absence of a hole(electron) in the filled Fermi sea is equivalent
   to the creation of an electron(hole), the above process can be viewed as emission into the conductors L and R of
   two electrons(holes) with opposite spins from the superconductor.

   We first consider a simple
   case in which the  transparencies $T_{L/R}$ of the tunnel barriers for spin-up and spin-down particles
   are the same.
   In the tunneling
   limit($T_{L,R} \ll 1$),  the outgoing
   state will be a superposition of the vacuum state $|0\rangle$,
   the local entangled state $(\mid \uparrow(-E)\rangle_L \mid \downarrow(E)\rangle_L-
   \mid \downarrow(-E)\rangle_L \mid
   \uparrow(E)\rangle_L)/\sqrt{2}$, and the nonlocal
   entangled state $(\mid \uparrow(-E)\rangle_L \mid \downarrow(E)\rangle_R-
   \mid \downarrow(-E)\rangle_L \mid
   \uparrow(E)\rangle_R)/\sqrt{2}$ with weight $\sqrt{1-T_L^2-T_LT_R\lambda(d)}$,
    $T_L, \sqrt{T_LT_R\lambda(d)}$, where
    $\lambda(d)\propto e^{-2d/\pi\xi}$ describes distance suppression
    of  breaking-up of a pair in  s-wave superconductors\cite{Recher}.

    \begin{figure}[t]
\epsfig{file=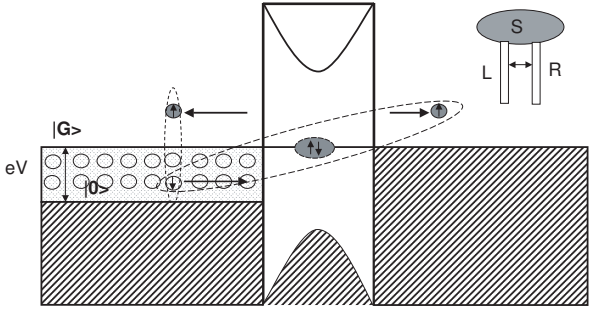, width=8cm}
 \caption{A schematic picture of direct and crossing Andreev reflection in the standard semiconductor model: two normal conductors
 (L and R) are connected to a superconductor(middle) via tunnel
 barriers, with a negative voltage $-eV$ being applied at the
 conductor L. The shaded region is the filled electron state. A hole(open circle) with spin down(up) below
 the superconducting chemical potential
 incident on the superconductor from L will be either converted as an electron(filled circle) with spin up(down)
 above the chemical potential at L or at R. The process can be viewed as  emission of an electron pair
 from the superconductor. The inset is the schematics of a hybrid structure with a superconductor and two normal conductors
 with separation $d$. }
\end{figure}

      Now we consider a general case allowing for an arbitrary value
      of barrier transparency and spin-flipping at the $NS$
      interface. Two normal conductors are labeled  by $L$
      and $R$ for brevity. A negative voltage $-eV$ is applied to
      the normal conductor $L$ to create entangled pairs of
      electrons.  Then the input state describing a stream of
      holes at energies $0<E<eV$ injecting to the superconductor can be written as
      \begin{equation}
      |\Psi_{in}\rangle=\prod_{0<E<eV}a_{L\uparrow}^{h\dagger}
      (E)a_{L\downarrow}^{h\dagger}(E)|G\rangle,
      \end{equation}
      where the ground state $|G\rangle$ is the filled Fermi sea up to the superconducting chemical potential
      in the normal conductors.
      $a_{X\sigma}^{\alpha\dagger}(E)$  is
      the creation operator exciting an incident spin-$\sigma$$(=\uparrow,\downarrow)$ hole($\alpha=h$)
      or electron($\alpha=e$) state
      with energy $E$(which is counted from the superconducting chemical potential) in normal conductor
      $X=L,R$, and similarly for operators $b$ of an outgoing state.

      Since in this work we want to  discuss the entanglement arising only from  Andreev reflection,
      a more general discussion in the presence of cotunneling events will be
      given elsewhere.  The experimental setup should satisfy the condition  $eV\ll \Delta$
      (the superconductor energy gap) where the cotunneling rate between the two normal conductors is
      negligibly small.
      Taking into consideration the crossed Andreev reflection, a scattering matrix relating
      the output state to the input state will be
      \begin{equation}
      \label{scattering matrix}
      \left(\matrix{b_L^e\cr b_R^e\cr b_L^h\cr
      b_R^h}\right)=\left(\matrix{S_{LL}^{ee}&0&S_{LL}^{eh}&S_{LR}^{eh}\cr 0&s_{RR}^{ee}&S_{RL}^{eh}&S_{RR}^{eh}\cr
      S_{LL}^{he}&S_{LR}^{he}&S_{LL}^{hh}&0\cr
      S_{RL}^{he}&S_{RR}^{he}&0&S_{RR}^{hh}}\right)\left(\matrix{a_L^e\cr a_R^e\cr a_L^h\cr
      a_R^h}\right),
      \end{equation}
      where $b_X^\alpha=(b_{X\uparrow}^\alpha,
      b_{X\downarrow}^\alpha)^T$, $a_X^\alpha=(a_{X\uparrow}^\alpha,
      a_{X\downarrow}^\alpha)^T$, with $T$ denoting the matrix transpose.
      The $2\times2$ submatrices
      $S_{XY}^{\alpha\beta}$  in Eq. $(\ref{scattering matrix})$ is spanned in the spin
      space, and the elements of the submatrices are
      $S_{XY;\sigma_1\sigma_2}^{\alpha\beta}$. They describe normal reflection, direct Andreev reflection and
      crossed Andreev reflection processes. The scattering matrix
      can be obtained from the scattering matrix approach\cite{Datta} by matching the wavefunctions at the $NS$ interface.
      In general, the
      scattering matrix has a pretty complicated relation to the elements of the scattering matrices of the barriers
      in the presence of crossed Andreev reflection.
      Fortunately, it takes a simple form in the tunneling limit, as
      we show later.

     Following Beenakker et al.\cite{Beenakker}, we rewrite the input state
     in the following form
      \begin{equation}
      \label{instate}
      |\Psi_{in}\rangle=\frac{i}{2}\prod_{0<E<eV}a_L^{h\dagger}\sigma_ya_L^{h\dagger
      T}|G\rangle.
      \end{equation}
       In Eq. $(\ref{instate})$, $\sigma_y$ is the  Pauli matrix
      $\left(\matrix{0&-i\cr i&0}\right)$. One finds from Eq.
      $(2)$
      \begin{equation}
      \label{a-b}
      a_L^{h\dagger}=b_L^{h\dagger}S_{LL}^{hh}+b_L^{e\dagger}S_{LL}^{eh}+b_R^{e\dagger}S_{RL}^{eh}.
      \end{equation}

         The input state evolves unitarily into the output state
      $|\Psi_{out}\rangle=U|\Psi_{in}\rangle=\prod_{i}Ua_i^\dagger U^\dagger |G\rangle$ in the
      Schr$\ddot{o}$dinger picture. However, in the Heisenberg picture, state vectors remain unchanged, and operators transform as
$a_i\leftrightarrow b_i=U^\dagger a_i U=\sum_j S_{ij}a_j$. The
operation $Ua_i^\dagger U^\dagger$ appearing in the evolution from
the input state to the output state
 can be considered as a time-reversal evolution as compared to the original transformation between $a_i$'s and $b_i$'s.
 Such an evolution is equivalent to  exchanging the role of inputs and
outputs, i.e., $U\Rightarrow U^\dagger$,  $S\Rightarrow S^\dagger$
and $a_i\Rightarrow b_i$. Then we have $Ua_i^\dagger
U^\dagger\Rightarrow U^\dagger b_i^\dagger U=\sum_j
S_{ji}b_j^\dagger$, which means that a direct substitution of
$b_i^\dagger$'s for $a_i^\dagger$'s  in the input state via the
related scattering matrix relation gives the correct output state.
Therefore, we obtain  the output state by inserting Eq.(\ref{a-b})
into Eq.(\ref{instate})
      \begin{eqnarray}
      \label{state out}
      |\Psi_{out}\rangle=i\prod_{0<E<eV}\Big\{(S_{LL}^{h h}\sigma_yS_{LL}^{h hT})_{12}b_{L\uparrow}^{h\dagger}
       b_{L\downarrow}^{h\dagger}+b_L^{e\dagger}S_{LL}^{eh}\sigma_y \nonumber\\S_{RL}^{ehT}b_R^{e\dagger
      T}+
      \sum_{X=L,R}\big[(S_{XL}^{eh}\sigma_yS_{XL}^{ehT})_{12}b_{X\uparrow}^{e\dagger}
      b_{X\downarrow}^{e\dagger}\nonumber\\
     +
      b_X^{e\dagger}S_{XL}^{eh}\sigma_y S_{LL}^{hhT}b_L^{h\dagger T}\big]
      \Big\}|G\rangle.
      \end{eqnarray}
     This expression reveals distinct physical processes.
      The first term represents normal reflection with the incoming holes reflecting backward at the NS interface, the second and third
      terms describe high-order Andreev reflection processes with the incoming holes transforming into electrons at the left and/or  right conductors,
      and the last term gives direct and crossed Andreev reflection, which
      constitutes the desired local and nonlocal entangled pairs of particles.

      Defining some auxiliary states as
      \begin{eqnarray}
      \label{auxiliary states}
|\psi_X\rangle&=&\frac{1}{\mathcal{A}_X}\prod_{0<\varepsilon<eV}b_X^{e\dagger}S_{XL}^{eh}\sigma_y
S_{LL}^{hhT}b_L^{h\dagger T}|G\rangle,\nonumber\\
|\psi_{LR}\rangle&=&\frac{1}{\mathcal{A}}\prod_{0<\varepsilon<eV}b_L^{e\dagger}S_{LL}^{eh}\sigma_y
S_{RL}^{ehT}b_R^{e\dagger T}|G\rangle, \nonumber
\end{eqnarray}
we rewrite the output state in the following form
 \begin{eqnarray}
      \label{state out intermediate}
      |\Psi_{out}\rangle&=&i\Big\{\mathcal{A}|\psi_{LR}\rangle+\mathcal{A}_X
      |\psi_X\rangle+
      \prod_{0<E<eV}(S_{LL}^{h h}\sigma_yS_{LL}^{h hT})_{12}\nonumber \\ && b_{L\uparrow}^{h\dagger}
       b_{L\downarrow}^{h\dagger}+\sum_{X=L,R}(S_{XL}^{eh}\sigma_yS_{XL}^{ehT})_{12}b_{X\uparrow}^{e\dagger}
      b_{X\downarrow}^{e\dagger}\Big\}.\nonumber
\end{eqnarray}
With the help of unitary conditions of the scattering matrix
(\ref{scattering matrix}), the state normalization condition
yields :
$\mathcal{A}_X=(Tr\Upsilon_{XL}\Upsilon_{LL}^\dagger)^{1/2}$,$\mathcal{A}=(Tr\Upsilon\Upsilon^\dagger)^{1/2}$,
with $\Upsilon_{XL}=S^{eh}_{XL}\sigma_y S^{hhT}_{LL}\sigma_y$,
$\Upsilon=S^{eh}_{LL}\sigma_y S^{ehT}_{RL}\sigma_y$.

          To manifest the entanglement more clearly, we transform from the electron-hole picture to
     an all-electron picture with the particle-hole transformation,  $
     c_{L\sigma}^{e\dagger}(-E)=b_{L\sigma}^{h}(E)$, $
     c_{L/R\sigma}^{e\dagger}(E)=b_{L/R\sigma}^{e\dagger}(E)$. The new vacuum
     state is defined by $|0\rangle=
     \prod_{0<E<eV;\sigma}a_{L\sigma}^{h\dagger}(E)
     |G\rangle$, and the re-normalized output
      state becomes to leading order in the Andreev reflection
      matrix $S^{eh}$
      \begin{eqnarray}
     \label{entangstate1}
      |\Psi_{out}\rangle&\simeq&
      \sqrt{1-\zeta_L-\zeta_R}|0\rangle+\sqrt{\zeta_L}|\psi_L\rangle+\sqrt{\zeta_R}|\psi_R\rangle,\nonumber\\
      |\psi_L\rangle&=&\prod_{0<E<eV}\zeta_L^{-1/2}c_L^{e\dagger}(E)\Upsilon_{LL}c_L^{e\dagger}(-E)|0\rangle,\nonumber\\
      |\psi_R\rangle&=&\prod_{0<E<eV}\zeta_R^{-1/2}c_R^{e\dagger}(E)\Upsilon_{RL}c_L^{e\dagger}(-E)|0\rangle,
      \end{eqnarray}
      where $\zeta_X=\mathcal{A}^2_X$. The output
      state is a superposition of the new defined vacuum state $|0\rangle$,
        and local and nonlocal entangled states $|\psi_L\rangle$,
      $|\psi_R\rangle$. The local entangled state consists of electron-electron pairs at the same
      conductors and nonlocal entangled states of electron-electron pairs at two different conductors.
       Notice that the entanglement states here are
      a product of two-particle entangled states at different
      energies.

      In the tunneling limit $T_{L/R\uparrow/\downarrow}\ll 1$,
      all the elements of the Andreev reflection matrices
      $(S^{eh}_{XL})_{ij}\ll 1$. Expanding the product in the output state
      (\ref{entangstate1}) to first order in $(S^{eh}_{XL})_{ij}$
      and using the fermion commutation relations
      repeatedly\cite{Samuelsson},
      the output state after rearranging operators becomes
       \begin{eqnarray}
       \label{entangstate2}
      |\Psi_{out}\rangle&\simeq&
      \sqrt{1-\zeta_L-\zeta_R}|0\rangle+\sqrt{\zeta_L}|\psi'_L\rangle+\sqrt{\zeta_R}|\psi'_R\rangle,\nonumber\\
      |\psi'_L\rangle&=&\int_0^{eV}dE\zeta_L^{-1/2}c_L^{e\dagger}(E)\Upsilon_{LL}c_L^{e\dagger}(-E)|0\rangle,\nonumber\\
      |\psi'_R\rangle&=&\int_{-eV}^{eV}dE\zeta_R^{-1/2}c_R^{e\dagger}(E)\Upsilon_{RL}c_L^{e\dagger}(-E)|0\rangle.
      \end{eqnarray}
      Such a state is a sum of superposition states of the newly defined
      vacuum state, and local and nonlocal entangled states at
      different energies, which  describes a wave-packet-like
      entanglement state.

      For a two-qubit pure state $|\psi\rangle$, one can quantify
      the degree of entanglement  by concurrence\cite{Wooters}
      $C=|\langle\psi|\sigma_y\otimes\sigma_y|\psi^*\rangle|$,
      where the superscript $*$ denotes complex conjugate.
      For the local and nonlocal entangled states $\psi_L$'s and $\psi_R$'s, one finds
      \begin{equation}
      \label{concurrence}
      C(\psi_{L/R})=2\frac{\sqrt{R_{L\uparrow}R_{L\downarrow}A_{LL/R\uparrow}A_{LL/R\downarrow}}}{R_{L\uparrow}
       A_{LL/R\downarrow}+R_{L\downarrow}A_{LL/R\uparrow}},
       \end{equation}
       which becomes in the tunneling limit($A_{LL/R\uparrow/\downarrow}\ll
       1$, since $T_{L/R\uparrow/\downarrow}\ll 1$)
       \begin{equation}
       \label{concurrence-tunneling}
         C(\psi_{L/R})=2\frac{\sqrt{A_{LL/R\uparrow}A_{LL/R\downarrow}}}
       {A_{LL/R\uparrow}+A_{LL/R\downarrow}},
       \end{equation}
     where $R_{L\uparrow}$, $R_{L\downarrow}$, $A_{LL/R\downarrow}$, and $A_{LL/R\uparrow}$ are the eigenvalues of
     the normal reflection matrix product $S_{LL}^{hh\dagger}S_{LL}^{hh}$  and the Andreev
      reflection matrix $S_{LL/R}^{eh\dagger}S_{LL/R}^{eh}$,
      respectively. In the absence of multiple Andreev
      reflections,  we find simple expressions for the eigenvalues of the normal and
      crossed Andreev reflection matrices: $A_{LL\uparrow/\downarrow}\approx T^2_{L\uparrow/\downarrow}/4$,
     $A_{LR\uparrow/\downarrow}\approx \lambda(d)
     T_{L\uparrow/\downarrow}T_{R\uparrow/\downarrow}/4$. Maximal
    entanglement is achieved in the spin-independent tunneling
    case. We  emphasize  that, the distance suppression of
crossed Andreev reflection does not degrade the degree of
entanglement. Its effect is just to diminish the weight of the
nonlocal entangled state $|\psi_R\rangle$.

     Experimentally, the degree of entanglement can be detected by violation of a Bell
     inequality\cite{Bell}. In solid state systems, the
     Bell inequality can be constructed in terms of the
     spin  correlators\cite{Lesovik,
Samuelsson, Beenakker}
    \begin{eqnarray}
    \label{BI}
    B&=&|F(\theta_X,\theta_Y)+F(\theta'_X,\theta_Y)+F(\theta_X,\theta'_Y)-F(\theta'_X,\theta'_Y)|\nonumber\\
     &\leq& 2.
    \end{eqnarray}
    In Eq. (\ref{BI}), the spin correlators are expressed in terms of
    low-frequency spin current correlators
    \begin{equation}
    F(\theta_X,\theta_Y)=\frac{\mathcal{N}_{\uparrow\uparrow}^{XY}-
    \mathcal{N}_{\uparrow\downarrow}^{XY}-\mathcal{N}_{\downarrow\uparrow}^{XY}+\mathcal{N}_{\downarrow\downarrow}^{XY}}
    {\mathcal{N}_{\uparrow\uparrow}^{XY}+
    \mathcal{N}_{\uparrow\downarrow}^{XY}+\mathcal{N}_{\downarrow\uparrow}^{XY}+\mathcal{N}_{\downarrow\downarrow}^{XY}},
    \end{equation}
where the low-frequency spin current correlators\cite{Buttiker}
are obtained as $\mathcal{N}_{\sigma_1\sigma_2}^{XY}=\int
d(t-t')<\delta I_{X\sigma_1}^h(t)\delta
I_{Y\sigma_2}^e(t')>=e^3V|(S^{hh}_{XX}S^{eh}_{XY})_{\sigma_1\sigma_2}|^2\times
x/h $  ($x=1$ for $X\neq Y$ and $x=2$ for $X=Y$). The spin current
correlators provide information about the joint probabilities to
find one spin-$\sigma_1$ particle at $X$ along the direction
$\theta_X$, and the other with spin-$\sigma_2$ at $Y$  along the
direction $\theta_Y$\cite{Samuelsson}. The direction along which
the detection is performed can be incorporated  by placing  a
unitary rotation matrix
$U_{\theta_X}=\left(\matrix{\cos\theta_X&-\sin\theta_X\cr
\sin\theta_X&\cos\theta_X}\right)$ before the normal reflection
and the Andreev reflection matrices\cite{Beenakker}. We find after
some algebra
\begin{equation}
\mathcal{N}_{\sigma_1\sigma_2}^{XY}=\frac{Tr
U^\dagger_{\theta_X}\sigma_zU_{\theta_X}S^{hh}_{XX}
 S^{eh\dagger}_{XY}U^\dagger_{\theta_Y}\sigma_zU_{\theta_Y}S^{eh}_{XY}
 S^{hh\dagger}_{XX}}{TrS^{hh\dagger}_{XX}S^{hh}_{XX}S^{eh\dagger}_{XY}S^{eh}_{XY}},
\end{equation}
where $\sigma_z=\left(\matrix{1& 0\cr 0&-1}\right)$. In order to
express the Bell-CHSH parameter $B$ in terms of the eigenvalues
$R_{L\sigma}$, $A_{LL/R\sigma}$ as well, we perform the polar
decomposition\cite{Mello} of the scattering submatrices with these
$2\times 2$ unitary matrices $U_1$, $U_2$, $U_3$, $V$:
$S_{LL}^{hh}=U_1R_L V^\dagger$, $S_{LL}^{eh}=U_2A_{LL} V^\dagger$,
$S_{LR}^{eh}=U_3 A_{LR}V^\dagger$, where the diagonal matrices are
$R_L=diag(R_{L\uparrow}^{1/2},R_{L\downarrow}^{1/2})$,
$A_{LL}=diag(A_{LL\uparrow}^{1/2},A_{LL\downarrow}^{1/2})$,
$A_{LR}=diag(A_{LR\uparrow}^{1/2},A_{LR\downarrow}^{1/2})$.
Performing a similar calculation as in Refs. \cite{Popescu} and
\cite{Beenakker} results in the maximal Bell-CHSH parameter $B$
for the local and nonlocal entangled spin states $|\psi_L\rangle$
and $|\psi_R\rangle$
\begin{equation}
\label{bell1}
B(\psi_{L/R})=2\Big[1+\frac{4R_{L\uparrow}R_{L\downarrow}A_{LL/R\uparrow}A_{LL/R\downarrow}}{(R_{L\uparrow}
       A_{LL/R\uparrow}+R_{L\downarrow}A_{LL/R\downarrow})^2}\Big]^{1/2}.
\end{equation}
In the tunneling limit, the maximal Bell-CHSH parameter becomes
\begin{equation}
\label{bell2}
 B(\psi_{L/R})=2\Big[1+\frac{4A_{LL/R\uparrow}A_{LL/R\downarrow}}{(
       A_{LL/R\uparrow}+A_{LL/R\downarrow})^2}\Big]^{1/2}.
\end{equation}
From Eq. (\ref{bell2}) one can readily obtain the following
expected relation $B=2(1+C)^{1/2}$ as compared with  Eq.
(\ref{concurrence-tunneling}).  The maximal Bell-CHSH paramter
always violates the Bell inequality as long as $R_{L\uparrow}$,
$R_{L\downarrow}$, $A_{LL/R\downarrow}$, $A_{LL/R\uparrow}\neq 0$.
In the spin-independent tunneling case, $B=2\sqrt{2}$, the Bell
inequality is maximally violated. As expected, there exists a
one-to-one correspondence between the concurrence $C$ and the
maximal Bell-CHSH parameter $B$ in the tunneling limit. We notice
that, the expressions of the concurrence and the maximal Bell-CHSH
parameter bear formal similarities compared to that in tunneling
between normal conductors\cite{Beenakker}.

The dephasing effect on the entanglement in our system can be
discussed in a way similar to Refs. \cite{Beenakker} and
\cite{Samuelsson}. Here we discuss possible experimental
realizations. For pure superconductors, the coherence length is of
order
 $\xi\sim \hbar/\delta p\sim\hbar v_F/ \Delta \sim
\frac{1}{ k_F} \frac{\epsilon_F}{\Delta} $.  $\epsilon_F$ is
typically $10^3\sim 10^4$ times $\Delta$, $k_F$ is of order of
$10^8$ $cm^{-1}$, and $\xi$ is typically $10^3\sim 10^{4}$
$nm$\cite{Ashcroft}.  Therefore, the exponential suppression with
increasing distance between two tunneling barriers poses no severe
restriction to experimental setup.   A pair of tunneling barriers
has been already prepared with distance of the order $10$
nm\cite{Russo}.  Since $k_F \xi \gg 1$, the distance suppression
is dominated by a polynomial term $(k_Fd)^{-2}$ if the
superconductor considered is three-dimensional. One way to reduce
or exclude such a power-law decay is to replace the
three-dimensional superconductor by a lower-dimensional
one\cite{Recher}.  Since the Andreev reflection process involves
two quasiparticles with one particle energy above and the other
below the superconducting chemical potential, energy-resolving or
spin-filtering current noise detectors can be used to measure the
low-frequency current correlators\cite{Martin}. Alternatively, one
can also use a pair of spin-resolved edge channels in the quantum
Hall effect proposed by Beenakker et al.\cite{Beenakker} for our
purpose.

In summary, we consider a unique accessible structure in which
both local and nonlocal entangled spin states for both electrons
and holes can be produced at the same time. The concurrence and
the maximal Bell-CHSH parameter have a simple dependence on the
eigenvalues of the normal and Andreev reflection matrices, and
possess a one-to-one correspondence in the tunneling limit.
Possible experimental realization is also discussed.

Z.Y.Zeng was supported by the NSFC under Grant No. 10404010, the
Project-sponsored by SRF for ROCS, SEM and the Excellent talent
fund of Jiangxi Normal University.This work was also supported in
part by Korea Research Foundation Grant No. KRF-2003–070–C00020, a
FRG grant of NUS and the DSTA under Project Agreement POD0410553.
Z. Y. Zeng also acknowledge the hospitality of ICTP at Trieste,
Italy where part of this work was done.

\end{document}